\documentclass[twocolumn,prl]{revtex4}
\usepackage{amssymb}
\usepackage{graphicx}
\usepackage{dcolumn}
\usepackage{bm}
\usepackage{amsmath}

\begin{document}

\title{Echo spectroscopy of bulk Bogoliubov excitations in trapped Bose-Einstein condensates}

\author{E. Gershnabel}
\author{N. Katz}
\author{R. Ozeri}
\author{E. Rowen}
\author{J. Steinhauer}
\altaffiliation{Present address: Department of Physics, Technion -
Israel Institute of Technology, Technion City, Haifa 32000,
Israel}
\author{N. Davidson}
\affiliation{Department of Physics of Complex Systems,\\
Weizmann Institute of Science, Rehovot 76100, Israel}

%\address{Department of Physics of Complex Systems,\\
%Weizmann Institute of Science, Rehovot 76100, Israel}

\begin{abstract}

We propose and demonstrate an echo method to reduce the
inhomogeneous linewidth of Bogoliubov excitations, in a
harmonically-trapped Bose-Einstein condensate. Our proposal
includes the transfer of excitations with momentum $+q$ to $-q$
using a double two photon Bragg process, in which a substantial
reduction of the inhomogeneous broadening is calculated.
Furthermore, we predict an enhancement in the method's efficiency
for low momentum due to many-body effects. The echo can also be
implemented by using a four photon process, as is demonstrated
experimentally.
\end{abstract}

\maketitle

Bragg spectroscopy of a trapped Bose-Einstein condensate (BEC) has
recently revealed many of the BEC's intriguing bulk properties
such as global coherence \cite{Ketterle_Doppler}, verification of
the Bogoluibov excitation spectrum, superfluidity and the
superfluid critical velocity, and suppression of low momentum
excitations due to quantum correlations of the ground state
\cite{Inhomogeneous}\cite{ours}\cite{enhancement}. However,
whereas the resonance frequency and the frequency integral of the
dynamic structure factor $S(k,\omega)$ mainly reflect the bulk
properties of the BEC, the frequency width of the spectrum is
dominated by inhomogeneous broadening mechanisms \cite{LDA}. The
inhomogeneous broadening is due to Doppler broadening
\cite{Ketterle_Doppler} and the inhomogeneous mean field energy
that is often well-described within a local density approximation
(LDA). Even when the LDA picture is not complete, as in the recent
observation of radial modes within the condensate, the
inhomogeneous broadening (manifested as the envelope function of
the multi-mode spectrum) still dominates
\cite{radial_modes}\cite{Tozzo}. Overcoming the inhomogeneous
mechanisms (arising from the particular trap shape), opens the
possibility of studying the homogeneous broadening mechanisms,
that can reflect the intrinsic decoherence processes of the bulk
excitations, e.g. elastic collisions with the BEC \cite{Collis}.

In this letter we propose a novel echo spectroscopic method, which
reduces the inhomogeneous broadening of bulk Bogoliubov
excitations in a BEC (in analogy to the familiar echo in two level
atoms \cite{Bloch}). This method is based upon the transfer of
Bogoliubov excitations from momentum $+q$ to $-q$, by a degenerate
two-photon Bragg transition induced by an optical lattice, with
wavenumber $2q$. Using the Gross-Pitaevskii equation (GPE), we
calculate the line shape of the echo excitation spectrum and show
a substantial reduction of the inhomogeneous broadening. We also
show a surprising quantum enhancement of the efficiency of this
method for low momentum, in contrast with the familiar suppression
due to structure factor considerations \cite{enhancement}.
Finally, we demonstrate the echo concept experimentally using an
alternative scheme based on a four-photon Bragg transition. Such a
scheme allows us to create $+q$ excitations by a two-photon Bragg
process \cite{Inhomogeneous}\cite{ours} and then transfer them to
$-q$ excitations via a degenerate four-photon Bragg process, using
the same laser beams.
\begin{figure}[tb]
\begin{center}
\includegraphics[width=8cm]{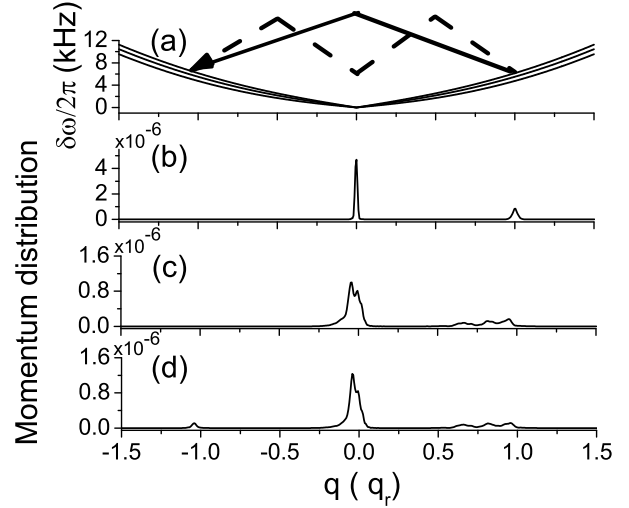}
\end{center}
\caption{The echo mechanism. (a) By application of the echo
standing wave ($\delta\omega\approx0$), $+q$ excitations are
transferred to $-q$ excitations. The echo process can be achieved
either by using a degenerate two-photon process (solid arrow) or a
degenerate four-photon process (dashed arrow). (b)-(d) Cross
sections of the wave function in longitudinal momentum space
during the various echo steps, for a momentum $q_{r}=8.06\mu
m^{-1}$, calculated with the GPE. The radial direction has been
integrated. (b) After a $1.3$ msec adiabatic Bragg pulse, (c)
after the Bragg pulse and $4.2$ msec delay. (d) As in (c), but
with an adiabatic $4.2$ msec two-photon echo pulse instead of the
delay.} \label{fig1}
\end{figure}

As shown in Fig. \ref{fig1}b, a short two-photon Bragg pulse with
wavenumber $+q$ initially excites the condensate uniformly
\cite{fourier}, along the $\hat{z}$ axis of the cylindrically
symmetric condensate. We then employ our echo scheme, shown
schematically in \ref{fig1}a. Specifically, the initial Bragg
pulse is followed by a second, much longer Bragg pulse with
wavenumber $2q$. For the frequency difference $\delta\omega \sim
0$ between the two Bragg beams, the echo resonance condition is
fulfilled for a transition from the positive momentum $q$ to the
negative momentum $-q$, as indicated in Fig. \ref{fig1}a by the
solid arrow and Fig. \ref{fig1}d. The various dispersion curves in
Fig. \ref{fig1}a are a schematic representation of the different
local Bogoliubov dispersion relations due to the inhomogeneous
local mean-field. They can also represent different radial modes
\cite{radial_modes}\cite{Tozzo}. The echo line-shape is expected
to be free of inhomogeneous effects, as long as there are no
transitions between the various curves, which belong to different
positions within the condensate (essentially within the LDA) or
different radial modes. Of course, the excitations do move along
$\hat{z}$ during the process and therefore every curve is coupled
with its vicinity, reflecting the fact that $q$ is not a good
quantum number in an inhomogeneous system of finite longitudinal
size. This combination of inhomogeneous mean field and finite
longitudinal size yields a residual small broadening of the
spectrum.

To verify the echo concept and calculate this residual broadening,
we use a simulation of the GPE \cite{simulation} and calculate
Bragg and echo processes for our experimental system
\cite{radial_modes}. Our experimental system contains $N=10^5$
atoms, confined in a harmonic trap with axial and radial angular
frequencies $w_{z}=2\pi \times 26.5$ Hz and $w_{r}=2\pi \times
226$ Hz, respectively, chemical potential $\mu/h=2.06$ kHz, and
LDA average healing length $\xi=0.23 \mu m$. All excitations are
axial, maintaining cylindrical symmetry \cite{radial_modes}.

\begin{figure}[tb]
\begin{center}
\includegraphics[width=8cm]{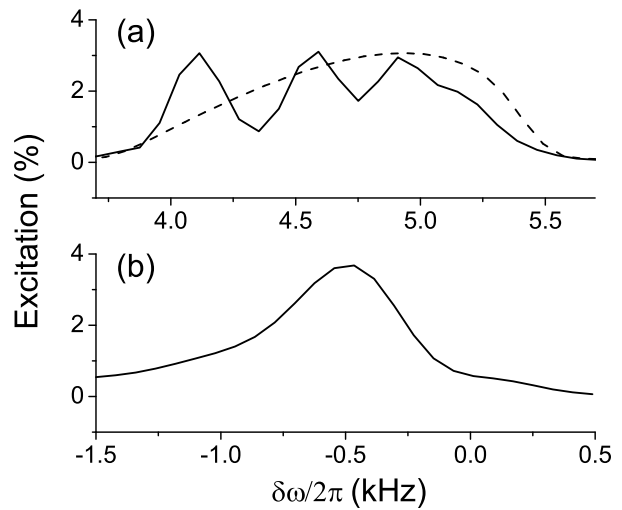}
\end{center}
\caption{Bragg and echo spectra for $q=8.06\mu m^{-1}$ (free
particle regime). (a) Excitation spectrum of the Bragg process
calculated with an exact GPE simulation (solid line) and in the
LDA (dashed line). (b) Excitation spectrum of the echo scheme from
a GPE simulation.} \label{fig2}
\end{figure}

Fig. \ref{fig1}b shows the calculated momentum distribution after
an initial $1.3$ msec Bragg pulse, where the $q=0$ ground state
component and $+q$ excitations are seen. The Bragg excitation
fraction is given as the integral of the wavefunction in momentum
space, divided by $\hbar q$. Fig. \ref{fig1}c shows the calculated
momentum distribution after an additional $4.2$ msec of free
evolution in the harmonic trap. The observed momentum components
smaller than $+q$ are the fraction of the excitations which have
left the condensate bulk during the free evolution and were
consequently slowed by the external harmonic trap. Fig.
\ref{fig2}d shows the calculated momentum distribution after an
initial $1.3$ msec Bragg pulse and $4.2$ msec echo pulse, where
the appearance of a $-q$ momentum population is seen. The echo
excitation fraction in Fig. \ref{fig1}d is obtained by subtracting
its total momentum from the total momentum of Fig. \ref{fig1}c,
and then dividing by $2\hbar q$. To suppress non-linear effects,
we adiabatically increase and decrease the intensity of our pulses
during the echo steps.

Using these GPE simulations, we first compare the spectrum of the
Bragg and echo excitations in the high-momentum, free-particle
regime shown in Fig. \ref{fig2} for $q_{r}=8.06\mu m^{-1}
(q\xi=1.85$). Fig. \ref{fig2}a shows the calculated Bragg spectrum
(using a $4.2$ msec rectangular pulse). The multi-peak structure,
corresponding to the recently observed radial modes
\cite{radial_modes} is evident in the spectrum. The calculated
FWHM of the spectrum is $1.26$ kHz. The dashed line in Fig.
\ref{fig2}a is the LDA line-shape \cite{LDA}. Although lacking the
multi peak structure, the LDA line-shape has a very similar width
to the exact GPE line-shape \cite{Tozzo}. Fourier broadening for a
$4.2$ msec rectangular Bragg pulse, Doppler broadening and the
collisional broadening (not included in the GPE calculations) are
$210.9$Hz, $135.5$Hz and $35.1$Hz (FWHM) respectively, all much
smaller than the inhomogeneous broadening \cite{comment_Doppler}.

Fig. \ref{fig2}b shows the calculated echo spectrum. A narrow peak
is seen close to $\delta\omega = 0$, as expected. Comparison with
Fig. \ref{fig2}a indicates that the echo has indeed reduced the
inhomogeneous linewidth. Thus, the echo pulse does not mix the
various radial modes. The echo FWHM is $0.59$ kHz, $2.1$ times
smaller then the Bragg FWHM of Fig. \ref{fig2}a. The Doppler
broadening, however, is still not resolved. The observed $\sim
 $0.5kHz negative shift of the resonance from $\delta\omega=0$ is
probably due to the decrease in momentum of the outgoing
excitations during the echo process (as explained above).

\begin{figure}[tb]
\begin{center}
\includegraphics[width=8cm]{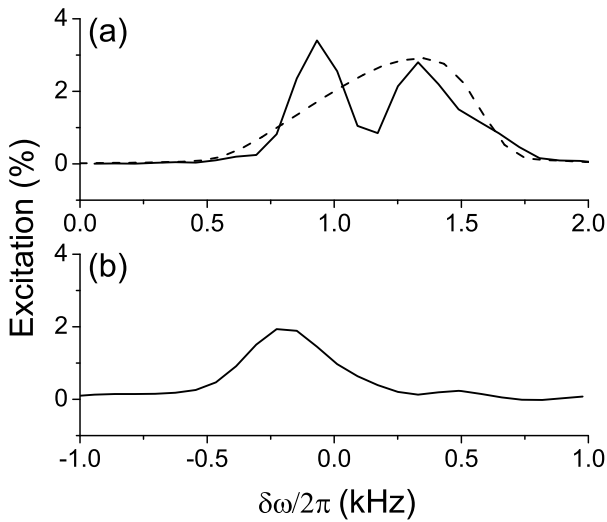}
\end{center}
\caption{Same GPE and LDA calculations as in Fig. \ref{fig2}, for
the Bragg (a) and echo (b) spectra, only with $q=3.10 \mu m^{-1}$
(phonon regime). We observe narrowing of the echo spectrum. In a
addition we note a smaller shift of the echo resonance, with
respect to fig. \ref{fig2}b.} \label{fig3}
\end{figure}

We repeat these calculations for low-momentum excitations in the
phonon regime with $q=3.10\mu m^{-1}$ ($q\xi=0.71$). The resulting
echo spectrum width (Fig. \ref{fig3}b) has a FWHM of $0.39$ kHz,
$1.7$ times smaller than the FWHM of the Bragg spectrum shown in
Fig. \ref{fig3}a. The shift of the echo resonance from zero, seen
in Fig. \ref{fig3}b, is smaller than the high-momentum case of
Fig. \ref{fig2}b, as expected.

We verify numerically that the width of the echo spectrum
increases for shorter and stronger echo pulses, indicating an
increase of Fourier and power broadenings, and decreases for
longer condensates (having the same $\mu$). Thus, having chosen
sufficiently weak and long pulses, the echo spectra of figures
\ref{fig2} and \ref{fig3} are finite-size broadened.

Next, we calculate the echo transition rate in the approximation
of an infinite, uniform Bogoliubov gas, and in analogy to the
calculation of the Bragg rate \cite{enhancement}. The initial
interaction Hamiltonian between the light and BEC is,
\begin{equation}
\label{hamiltonian}
\hat{H}^{'}=C\sum_{klmn}\hat{c}_{l}^{\dag}\hat{a}_{n}^{\dag}\hat{c}_{k}\hat{a}_{m}\delta_{l+n-k-m}
\end{equation}
Where $C$ is the coupling constant, $\hat{c}$
($\hat{c}^{\dagger}$) are the photonic annihilation (creation)
operators and $\hat a$ ($\hat a^{\dagger}$) are the atomic
annihilation (creation) operators. The first transition we analyze
is that from $+q$ to $-q$, i.e. from the initial state
$|i\rangle=|n_{q},n_{-q};N_{q},N_{-q}\rangle$ to the final state
$|f\rangle=|n_{q}+1,n_{-q}-1;N_{q}-1,N_{-q}+1\rangle$ where
$n_{q}$ represents the number of photons in the $+q$ direction and
$N_{q}$ represents the number of $+q$ Bogoliubov excitations.
Since the excitations are axial, we refer only to the magnitudes
of the momenta. In order to evaluate the transition matrix
element, we must consider the matrix element of the Hamiltonian
(\ref{hamiltonian}) between $|i\rangle$ and $|f\rangle$, and
transform the atomic operators into Bogoliubov excitation
operators. We find
\begin{equation}
\label{matrix_element} \langle
f|\hat{H}^{'}|i\rangle_{+}=C\sqrt{n_{-q}}\sqrt{n_{q}+1}\sqrt{N_{q}}\sqrt{N_{-q}+1}\{u_{q}^{2}+v_{q}^{2}\}
\end{equation}
where $u_{q}$ and $v_{q}$ are the Bogoliubov amplitudes
\cite{Fetter} and the "+" notation represents the transition from
$+q$ to $-q$. The second transition to be explored is the reverse
process (transfer from $-q$ to $+q$), whose matrix element is
denoted by $\langle f|\hat{H}^{'}|i\rangle_{-}$. In the Fermi
golden rule approximation the transition rate is given by
$Rate=\frac{2\pi}{\hbar}(|\langle
f|\hat{H}^{'}|i\rangle_{+}|^{2}-|\langle
f|\hat{H}^{'}|i\rangle_{-}|^{2})\delta (E_{q}-E_{-q})$. Assuming
the initial condition $N_{q}\gg 1$, i.e. $(N_{q}+1)\approx N_q$,
neglecting the $n_{q}N_{-q}$ term, and assuming a classical laser
field (i.e. $n_{q}\gg N_{-q}$), the rate can be approximated by
$Rate\approx\frac{2\pi}{\hbar}|C|^{2}N_{q}n_{-q}n_{q}\{u_{q}^{2}+v_{q}^{2}\}^{2}\delta
(E_{q}-E_{-q})$. Hence, there is no Bosonic amplification due to
the least populated mode ($N_{-q}$ in our case). Considering the
term $\{u_{q}^{2}+v_{q}^{2}\}^{2}$ and using the normalization
$u_{q}^2-v_{q}^2=1$ and the value of the structure factor
$S_{q}=(u_{q}-v_{q})^2$ we find that the rate per excitation with
momentum $q$ is proportional to the response $R$, defined by
$R\equiv \left(\frac{1+S_{q}^2}{2S_{q}}\right)^{2}$. $R$ has the
low $q$ asymptotic behavior of $(2S_{q})^{-2}$, thus for low $q$
(where $S_{q}$ is small) we calculate a large enhancement in $R$.
\begin{figure}[tb]
\begin{center}
\includegraphics[width=8cm]{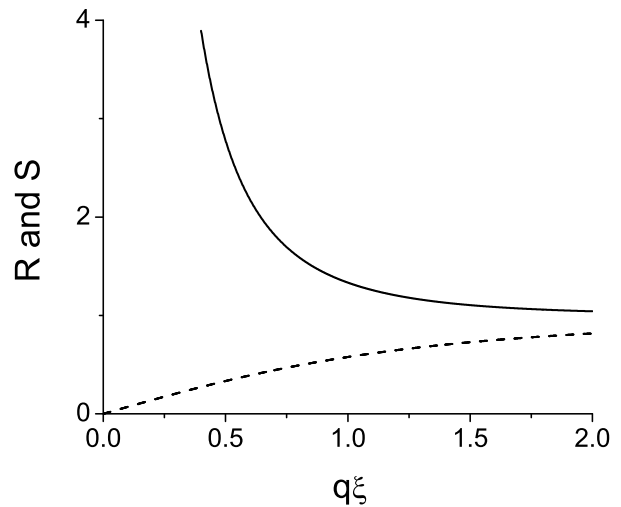}
\end{center}
\caption{The echo response $R$ (solid line). The dashed line shows
the structure factor $S_q$. Note the enhancement of $R$ and the
suppression of $S_q$ for small $q$.} \label{fig4}
\end{figure}

The response is seen in Fig. \ref{fig4} (solid line) to be greatly
enhanced for small $q$, in contrast with the Bragg process, which
is suppressed by the structure factor (dashed line). We emphasize
that the validity of these results is for the infinite homogeneous
gas. We do not observe such an enhancement in the GPE calculations
for a trapped BEC.

To demonstrate the echo spectroscopy experimentally, we use the
apparatus described in \cite{ours}, in which a nearly pure
($>90\%$) BEC of  $~1\times 10^{5}$ $^{87}$Rb atoms in the
$|F,m_{f}\rangle=|2,2\rangle$ ground state, is formed in a
cylindrically symmetric magnetic trap. A two-photon Bragg
transition and the subsequent four-photon echo transition are both
induced by the same counter-propagating (along $\hat{z}$) laser
beam pair, locked to a Fabri-Perot cavity line, detuned $44$GHz
below the $5S_{1/2},F=2\longrightarrow 5P_{3/2},F^{\prime }=3$
transition.

\begin{figure}[tb]
\begin{center}
\includegraphics[width=8cm]{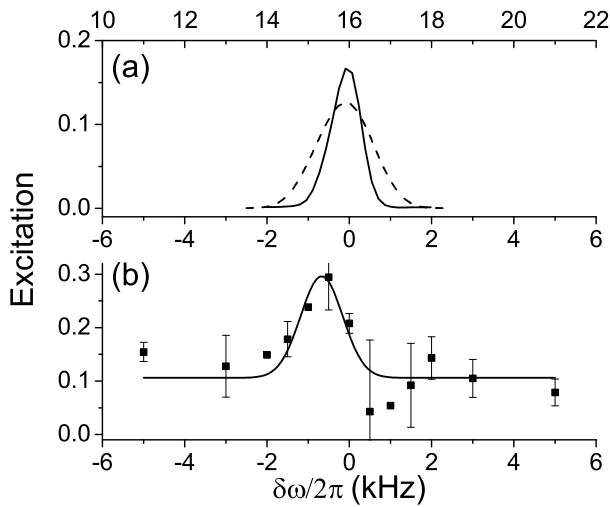}
\end{center}
\caption{(a) GPE predictions. The solid curve and the lower axis
show the four-photon echo spectrum for the smoothly varying $1$
msec pulse used in the experiment. The dashed curve and the upper
axis correspond to the Bragg spectrum for the same pulse. (b)
Experimental echo spectrum (squares) and a Gaussian fit (solid
line). } \label{fig5}
\end{figure}

After exciting approximately $35\%$ of the condensate by means of
a short $0.1$ msec rectangular Bragg pulse with $\delta \omega =
2\pi \times 16$ kHz, we apply a 1 msec echo pulse with an envelope
shaped as $\sin(\pi\times 10^{3}t)$ ($0<t<1$ msec), using the same
beams. We vary $\delta \omega$ to make a spectroscopic measurement
of the echo response around $\delta \omega=0$, using the
four-photon process to transfer excitations from $+q$ to $-q$
(Fig. \ref{fig1}a, dashed arrow). From the absorption images after
38 msec of time-of-flight we find the ratio between the $-q$
population $N_{-q}$ and the total (not collided) amount of atoms
$N_{q}+N_{BEC}+N_{-q}$. The measured echo spectrum is shown in
Fig. \ref{fig5}b together with a Gaussian fit to the data,
centered close to zero frequency, and with a FWHM of $1.21$ kHz.
The experimental echo duration is limited by collisions, which are
not taken into account in our GPE simulation and cause a
significant degradation in our signal for echo times longer than
$1$ msec. Sloshing of the condensate \cite{radial_modes} may also
cause a broadening of the measured spectrum width. However, we
still observe a sub-Fourier broadening of the echo spectrum, which
is also predicted by a GPE simulation (for four-photon echo), as
shown by the solid line in Fig. \ref{fig5}a (FWHM $0.86$ kHz). The
dashed line in Fig. \ref{fig5}a, corresponds to a Bragg spectrum
of a 1 msec pulse (with the same envelope as the echo), and is
Fourier limited (FWHM $1.63$ kHz). The sub-Fourier width of the
echo spectrum is possible due to the non-linearity of the
four-photon process. Thus, even with relatively short pulses, a
narrowing of the lineshape is achieved, as compared to standard
Bragg spectroscopy \cite{Inhomogeneous}\cite{ours}.

In conclusion, we implement a spectroscopic echo method in
momentum space in order to reduce the measured inhomogeneous
linewidth of Bogoliubov excitations in a trapped BEC. We note that
by varying the shape of the trap (either in length or in
functional form) the collisional and finite-size broadenings may
be reduced further than reported here. We also predict an
enhancement of the echo rate at low momentum emerging from
constructive interference of the amplitudes of various quantum
paths for this process. This is in contrast to the usual
suppression of low momentum processes in quantum degenerate Boson
systems.

This work was supported in part by the Israel Ministry of Science
and the Israel Science Foundation.


\begin{references}
\bibitem{fourier} The Bragg pulse must be short enough such that its
Fourier broadening dominates any other broadening mechanism to
excite the condensate uniformly.
\bibitem{simulation}  We numerically solve the GPE $i\hbar \partial _{t}
\psi=\{-\hbar^{2}\nabla^{2}/(2m)+V+g|\psi|^{2}\}\psi$, with
$V(\vec{r},t)=m/2(w_{r}^{2}r^{2}+w_{z}^{2}z^{2})+\Omega(t)V_{B}cos(qz-\delta
\omega t)$. $g$ is the mean-field coupling constant, $m$ is the
$^{87}$Rb atomic mass, $q$ is the wavenumber of the excitation,
$V_{B}$ is the potential intensity and $\Omega(t)$ is an envelope
function. Using cylindrical symmetry we evolve $\psi$ on a two
dimensional grid $N_{z}N_{r}($up to $4096\times 32)$, using the
Crank-Nicholson differencing method.
\bibitem{comment_Doppler} The Doppler line-shape prediction is $|\psi(p_{z})|^{2}\sim \{2(4+\kappa^{2})J_{1}(\kappa) J_{2}(\kappa)+\kappa J_{0}(\kappa)[5\kappa J_{1}(\kappa)-16J_{2}(\kappa)+3\kappa J_{3}(\kappa)]\}/\kappa^{3}$
\cite{Ketterle_Doppler}, where $\kappa\equiv p_{z}z_{0}/\hbar$,
$z_{0}$ and $p_{z}$ are the axial Thomas-Fermi radius and the
initial condensate momentum, respectively. We take
$\hbar\omega-\hbar\omega_{0}=p_{z}\cdot\hbar q/m$ where $\hbar q$
is the momentum given to the condensate and $\hbar\omega_{0}$ is
the free particle resonance energy. The collisional line-shape is
approximated by the Lorenzian
$1/((\omega-\omega_{0})^{2}/(2\pi)^2+(\Gamma /2)^{2})$ where
$2\pi\Gamma$ is the collision rate \cite{Collis}.
\end{references}
\end{document}